\documentclass[twocolumn,pre,showpacs]{revtex4}
\usepackage{graphicx}
\usepackage{epsfig}
\usepackage{bm}

\begin{document}

\title{Reaction, L\'{e}vy Flights, and Quenched Disorder}

\author{Ligang Chen}
\author{Michael W. Deem}

\affiliation{Department of Chemical Engineering\\
University of California\\
Los Angeles, CA  90095--1592}

\begin{abstract}
We consider the $A + A \to \emptyset$ reaction, where the transport of
the particles is given by L{\'e}vy flights in a quenched random
potential.  With a common literature model of the disorder, the random
potential can only increase the rate of reaction.  With a model of the
disorder that obeys detailed balance, however, the rate of reaction
initially increases and then decreases as a function of the disorder
strength.  The physical behavior obtained with this second model is in
accord with that for reactive turbulent flow, indicating that L{\'e}vy
flight statistics can model aspects of turbulent fluid transport.
\end{abstract}

\maketitle

\section{Introduction}
L\'{e}vy flights have been used to model a variety of physical processes
such as epidemic spreading~\cite{Mollison77}, self-diffusion in micelle
systems~\cite{Urbach}, and transport in heterogeneous
rocks~\cite{Shlesinger}. L\'{e}vy flights are essentially a generalization
of ordinary Brownian walks. The normalized step size distribution for
L\'{e}vy flights in $d$ dimensions is
\begin{equation}
P(r)d^dr = \frac{f r_0^f}{S_d} r^{-1-f} dr d\Omega
\label{1}
\end{equation}
where $r$ is the step size, $f$ is the step index, $S_d =
2\pi^{d/2}/(d/2-1)!$, and $r_0$ a lower microscopic step cutoff.  In
the case of $f=2$, we recover Brownian motion. However, for $f<2$, the
distribution of step sizes exhibits a long-range algebraic tail that
corresponds to large but infrequent steps, so called {\it rare
events}. Due to these rare, large steps, the mean square step size
deviation diverges, and the central limit theorem does not hold. The
rare, large step events prevail and determine the long time
behavior. The dynamic exponent $z$ that characterizes the mean
displacement as a function of time by $\langle \vert r(t) \vert \rangle^2
 \sim ({\rm const}) t^{2/z}$ depends on the microscopic step index $f$
according to the relationship $z=f$, indicating anomalous enhanced
diffusion, that is, superdiffusion.

It is well-known that quenched random disorder leads to sub-diffusive
behavior in two-dimensional Brownian walks. L\'{e}vy flights in such random
environments have attracted increasing attention recently. The
interplay between the ``built-in'' superdiffusive behavior of the
L\'{e}vy flights and the effect of the random environment generally
leading to sub-diffusive behavior has been
examined~\cite{Fogedby,Honkonen96,Honkonen00}. Surprisingly, 
an $\epsilon$ expansion shows that in the
models of random potential disorder examined to date, the dynamic exponent $z$
locks onto the L\'{e}vy flight index $f$ in any dimension,
regardless of the range or
strength of disorder (notwithstanding some claims 
regarding divergence-free disorder in \cite{Honkonen00}).

The behavior of chemical reactions with random
potential~\cite{Deem981,Deem982} and isotropic turbulence
disorder~\cite{Park98,Deem99} has been examined. Reactants diffusing
according to L\'{e}vy flight statistics have also been studied in a
model of branching and annihilating processes~\cite{Vernon2001}. In
general, potential disorder tends to slow down the diffusing
reactants. Since these reactions typically become transport-limited at
long times, potential disorder tends to slow down the reaction as
well. It was found, however, that a small amount of potential disorder
added to the turbulent fluid mixing leads to an increased rate of
reaction. This phenomenon of 'superfast' reaction occurs because the
disorder traps reactants in local potential wells, which quadratically
increases the local reaction rate, while the turbulence rapidly
replenishes the reacting species to these regions. As the potential
disorder is increased, eventually the rate of reaction decreases, due
to a slowing of the transport. It is interesting to study the behavior
of reactants following L\'{e}vy flight statistics in quenched random
disorder. The question is: Can the L\'{e}vy statistics mimic rapid
turbulent transport and so lead to superfast reaction? Furthermore, do
the reactions become transport limited or reaction limited at long
times?

In this paper, we analyze two different models of the disorder. A
conventional literature model, which does not satisfy detailed
balance, is discussed in section II.  A new model that does satisfy
detailed balance is discussed in section III. We study the effect of
L\'{e}vy flight statistics and quenched random disorder on the simple
bimolecular recombination reaction in two dimensions. 
We focus on the physically meaningful two-dimensional case
because the effects disappear above the upper critical dimension that
is near two and
because other, exact methods of analysis are probably more appropriate
in one dimension.
Detailed results
of the field theoretic renormalization are presented. We conclude this
paper in section IV.

\section{Reaction in a Common Literature Model of Disorder}
Including the normal diffusion term, the Fokker-Planck equation for
L\'{e}vy flights in a quenched potential field has been modeled
by~\cite{Fogedby,Honkonen96,Honkonen00}
\begin{eqnarray}
\frac{\partial G({\bf r}, t)}{\partial t} &=& D_2 \nabla^2 G({\bf r}, t)
+ D_f (\nabla^2)^{f/2} G({\bf r}, t) \nonumber\\ 
 & &+ \nabla \cdot[G({\bf r}, t)
\nabla V ({\bf r})]
\label{Fokker}
\end{eqnarray}
where $(\nabla^2)^{f/2}$ is interpreted as the inverse Fourier
transform of $-k^f$, which is a spatially non-local integral operator
reflecting the long-range character of the L\'{e}vy steps with
microscopic step index $f$. Fourier transforms are defined as
$\hat{f}({\bf k}) = \int d^d x e^{i{\bf k} \cdot {\bf x}} f({\bf
x})$. The last term on the right hand side of Eq.\ (\ref{Fokker})
is a drift term due to the
motion of the walker in the force field.  We assume a Gaussian
distribution of the random potential force field, $V({\bf r})$, with
correlation
\begin{equation}
\langle V({\bf k}_1) V({\bf k}_2) \rangle = \frac{\gamma}
{k^{2+y}} (2\pi)^d \delta({\bf k}_1+{\bf k}_2)
\label{VV}
\end{equation}
Note that the strength of disorder is parameterized by $\gamma$, whereas
the range is parameterized by $y$.

The reaction we are considering is
\begin{equation}
A+A
~{\mathrel{\mathop{\to}\limits^{\lambda}}}~ \emptyset
\end{equation}
A field theory is derived by identifying a master equation and using
the coherent state mapping~\cite{Lee}. The random potential is
incorporated with the replica trick~\cite{Yudson}. The action within
the field theory is
\begin{eqnarray}
S &=& \int d^d {\bf x} \int_0^{t_f} dt \bar{a}_i({\bf x}, t) [\partial_t
- D_f(\nabla^2)^{f/2} - D_2 \nabla^2 \nonumber\\ 
  & & + \delta(t)] a_i({\bf x}, t) 
+ \frac{\lambda}{2} \int d^d {\bf x}
\int_0^{t_f} dt [2 \bar{a}_i ({\bf x}, t) a_i^2({\bf x}, t)\nonumber\\ 
  & & + \bar{a}_i^2({\bf x}, t) a_i^2({\bf x}, t)] - n_0 \int d^d {\bf
x} \bar{a}_i({\bf x}, 0) \nonumber \\
  & & - \frac{\gamma}{2} \int dt_1 dt_2 \int_{{\bf k}_1 {\bf k}_2 {\bf
  k}_3{\bf k}_4} (2\pi)^d \delta({\bf k}_1 + {\bf k}_2 +  {\bf k}_3 +
  {\bf k}_4) \nonumber\\ 
  & &\times \frac{ {\bf k}_1 \cdot ({\bf k}_1 + {\bf k}_2) {\bf k}_3
  \cdot ({\bf k}_3 + {\bf k}_4)}{|{\bf k}_1 + {\bf k}_2|^{2+y}}
   \nonumber \\
  & & \times \hat{\bar{a}}_i ({\bf k}_1, t_1) \hat{a}_i ({\bf k}_2, t_1)
  \hat{\bar{a}}_j ({\bf k}_3, t_2) \hat{a}_j ({\bf k}_4,
  t_2)
\end{eqnarray}
Summation is implied over replica indices. The notation $\int_{\bf k}$
stands for $\int d^d{\bf k}/(2\pi)^d$.

The concentration of the reactant A at time $t$, averaged over the
initial conditions, is given by
\begin{equation}
c_A({\bf x},t) = \lim_{N \to 0} \langle a_i({\bf x},t) \rangle
\label{8}
\end{equation}
where the average is taken with respect to  $\exp(-S)$.

To apply the field theoretic renormalization procedure, the action is
recast~\cite{Justin} as
\begin{eqnarray}
S &=& \int d^d {\bf x} \int_0^{t_f} dt \bar{a}_i({\bf x}, t) [\partial_t
- Z_fD_{fR}(\nabla^2)^{f/2} \nonumber \\ 
  & &- \mu^{f-2} Z_2 D_{2R} D_{fR} \nabla^2 + \delta(t)]
a_i({\bf x}, t) \nonumber \\
  & & + \frac{1}{2} \mu^{f-d} Z_{\lambda} \lambda_R D_{fR} \int d^d
{\bf x} \int_0^{t_f} dt [2 \bar{a}_i({\bf x}, t) a_i^2({\bf x}, t) \nonumber\\ 
  & & + \bar{a}_i^2({\bf x}, t) a_i^2({\bf x}, t)] - \mu^d n_{0R} \int
d^d {\bf x} \bar{a}_i({\bf x}, 0)  \nonumber \\
  & & - \frac{1}{2} \mu^{\epsilon} Z_{\gamma} \gamma_{R} D_{fR}^2 \int
dt_1 dt_2 \int_{{\bf k}_1 {\bf k}_2 {\bf k}_3{\bf k}_4} (2\pi)^d \nonumber \\ 
  & &\times \delta({\bf k}_1 + {\bf k}_2 +  {\bf k}_3 + {\bf k}_4) \frac{
{\bf k}_1 \cdot ({\bf k}_1 + {\bf k}_2) {\bf k}_3 \cdot ({\bf k}_3 +
{\bf k}_4)}{|{\bf k}_1 + {\bf k}_2|^{2+y}} \nonumber \\
  & & \times \hat{\bar{a}}_i ({\bf k}_1, t_1) \hat{a}_i ({\bf k}_2, t_1)
  \hat{\bar{a}}_j ({\bf k}_3, t_2) \hat{a}_j ({\bf k}_4,
  t_2)
\end{eqnarray}
where the renormalization constants $Z_f, Z_2, Z_{\lambda}$, and
$Z_{\gamma}$ have been introduced to absorb the UV divergences of the
model. The parameters, $D_{fR}, D_{2R}, \lambda_R,$ and $\gamma_R$ are
the dimensionless expansion parameters of the model. Since the
L\'{e}vy flight term, $D_{fR}(\nabla^2)^{f/2}$, is the most important
term, it is chosen as the dimensionless free term, i.e. the propagator
is $[\partial_t - D_{fR}(\nabla^2)^{f/2}]^{-1}$. Note, we are not
allowed to treat the regular diffusion term as free term, as this
violates the physics of the scaling and leads to a diverging renormalized
$D_f$.  Now, the critical dimension following from standard
power counting of $\gamma_R$ is $d_c =
2f + y - 2$, and we introduce for an $\epsilon$ expansion $\epsilon =
d_c - d$. The scale-setting wavenumber parameter is denoted by $\mu$,
and we assign the dimensions to the rest of the terms accordingly.

The connections between the renormalized and unrenormalized parameters
are
\begin{eqnarray}
Z_fD_{fR} &=& D_f \nonumber \\
\mu^{f-2} Z_2 D_{2R} D_{fR} &=& D_2\nonumber\\
\mu^{f-d} Z_{\lambda} \lambda_{R} D_{fR} &=& \lambda\nonumber\\
 \mu^d n_{0R} &=& n_0\nonumber\\
\mu^{\epsilon} Z_{\gamma} \gamma_{R} D_{fR}^2 &=& \gamma
\end{eqnarray}

\begin{figure}[t]
\centering
\leavevmode
\includegraphics[height=2in, angle=0]{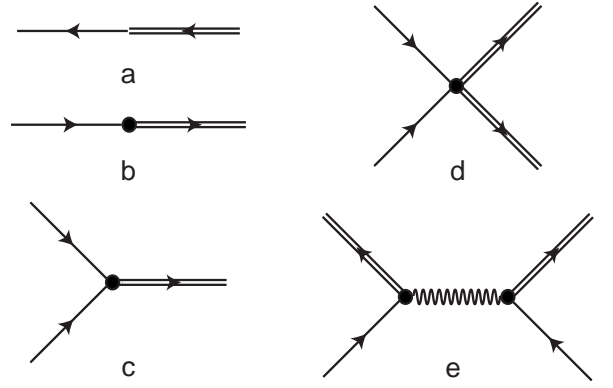}
\caption{(a) Diagram representing the propagator. The arrow points
in the direction of increasing time, and double lines represent the
bar fields. (b) Normal diffusion vertex
$D_2$. (c), (d) Reaction vertices $\lambda$. (e) Disorder
vertex $\gamma$.}
\label{fig1}
\end{figure}

\begin{figure}[t]
\centering
\leavevmode
\includegraphics[height=5in, angle=0]{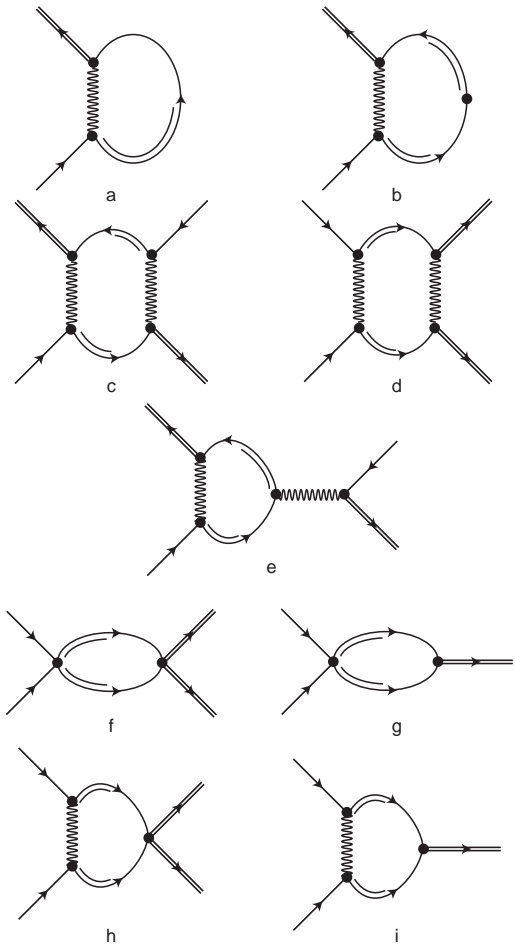}
\caption{One-loop diagrams: (a) self-energy diagram contributing to
$D_f$; (b) self-energy diagram contributing to $D_2$; (c), (d), and
(e) vertex diagrams contributing to $\gamma$; (f), (g), (h), and (i)
vertex diagrams contributing to $\lambda$.}
\label{fig2}
\end{figure}

To one-loop order, self-energy diagrams and vertex diagrams are
summarized in Fig.~\ref{fig1} and Fig.~\ref{fig2}. We may be tempted
to use the momentum-shell renormalization procedure here. However, due
to the difficulty of regularizing this action, the first self-energy
diagram would incorrectly contribute to $D_{2R}$ by the momentum-shell
renormalization, rather than to $D_{fR}$ by the field theoretic
renormalization that is consistent with perturbation theory.
In the evaluation of the diagram in Fig.~\ref{fig2}a, it is
important to treat the external momentum exactly.  If a series
expansion in the external momentum is performed on the integrand, rather than
on the result of the integral, an incorrect contribution to $Z_2$ arises.
Interestingly, when the diagram in Fig.~\ref{fig2}a is evaluated, only
a finite contribution to $Z_f$ results.
Complete calculation shows
\begin{eqnarray}
Z_f &=& 1\nonumber\\
Z_{2} &=& 1 - \frac{(2f-3)\gamma}{4\pi \epsilon}\nonumber\\
Z_{\lambda} &=& 1 - \frac{\gamma_R}{2\pi \epsilon} +
\frac{\lambda_R} {4\pi (f-d)}\nonumber\\
Z_{\gamma} &=& 1 + \frac{\gamma_R}{2\pi \epsilon}
\end{eqnarray}
where a double pole expansion, $1/\epsilon$ and $1/(f-d)$, is used in
the calculation of $Z_{\lambda}$. The use of the double pole expansion
in two dimensions
means we consider $f$ just slightly smaller than $2$ and $y$ small.
 As usual, the $\beta$ functions defined by
\begin{eqnarray}
\beta_{D_{2R}} &=& \mu \frac{\partial}{\partial \mu} D_{2R}\nonumber\\
\beta_{\lambda_R} &=& \mu \frac{\partial}{\partial \mu} \lambda_R\nonumber\\
\beta_{\gamma_R} &=& \mu \frac{\partial}{\partial \mu} \gamma_R
\end{eqnarray}
give the flow equations in two dimensions as
\begin{eqnarray}
\frac{dD_{2R}}{dl} &=& -\beta_{D_{2R}} = (f-2) D_{2R} + \frac{(2f-3)
\gamma_R D_{2R}}{4\pi}\nonumber\\
\frac{d\lambda_R}{dl} &=& -\beta_{\lambda_R} = (f-2)\lambda_R +
\frac{\gamma_R \lambda_{R}}{2\pi} - \frac{\lambda_R^2}{4\pi}\nonumber\\
\frac{d\gamma_R}{dl} &=& -\beta_{\gamma_R} = \epsilon \gamma_R -
\frac{\gamma_R^2}{2\pi}
\end{eqnarray}
where we use the relation $\mu = \Lambda/e^l$, and $\Lambda$ is a
microscopic cutoff. Since $Z_f = 1$, the reaction and disorder terms
do not affect the dynamic exponent, and $z = f$.
 We determine the long-time decay from the flow equation
via matching to short-time perturbation theory. The flow equations are
integrated to a short time such that
\begin{equation}
t(l^*) = t \exp[-\int_0^{l^*} z(l)dl] = t_0
\end{equation}
At short times, we find the mean displacement of an unreactive
particle from $\langle \vert r(t(l^*), l^*) \vert \rangle^f =
 4 D_f t(l^*)$, and the
concentration of reactants from $c_A(t(l^*), l^*) = [(n_{0R}(l^*)
\Lambda^d)^{-1} + D_f \lambda_R(l^*) \Lambda^{f-2} t(l^*)]^{-1}$. The
long-time asymptotic values are given by scaling: $\langle
\vert r(t) \vert \rangle^f =
e^{f l^*} \langle \vert r(t(l^*), l^*) \vert \rangle^f$, and
$c_A(t) = e^{-2l^*} c_A(t(l^*), l^*)$.

We first investigate the behavior of $D_{2R}$.
As we will see below, the fixed point for
$\gamma_R$ is $\max(0, 2 \pi \epsilon)$.  Using this result,
we see that $D_{2R}$ flows exponentially to zero
as long as $y < 4(2-f)$, when $f$ is near 2.
Likely, a higher loop calculation would extend the region in which
$D_{2R}$ flows to zero.
Thus, at least within the region in which our flow equations apply,
$D_{2R}$ always flows to 0 in the presence
of L\'{e}vy flights. It is, therefore, unnecessary to introduce such a
normal diffusion term in the model.

\begin{figure}[t]
\centering 
\leavevmode 
\includegraphics[height=2in, angle=-90]{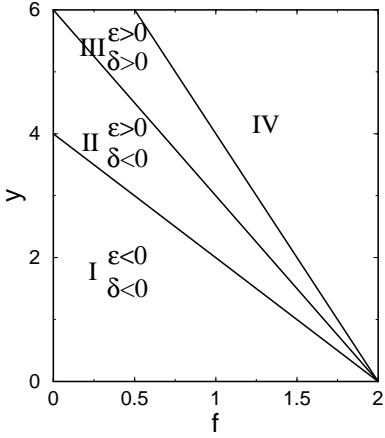}
\caption{Different regions predicted by the flow equations for the
L\'{e}vy flight system with disorder model I in two dimensions.
The flow equation are
accurate for small $y$ and $f$ slightly less than 2.  The flow
equations do not apply in region IV.
}
\label{fig3}
\end{figure}

For $\epsilon < 0$, i.e. in region I of Fig.~\ref{fig3}, there is
only a set of trivial stable fixed points, $\gamma_R^* = 0$ and
$\lambda_R^*=0$, for the system. The matching procedure gives the
normal concentration decay as
\begin{equation}
c_A(t) \sim \frac{1}{\lambda t}
\end{equation}

\begin{figure}[tb]
\centering
\leavevmode
\includegraphics[height=2in, angle=-90]{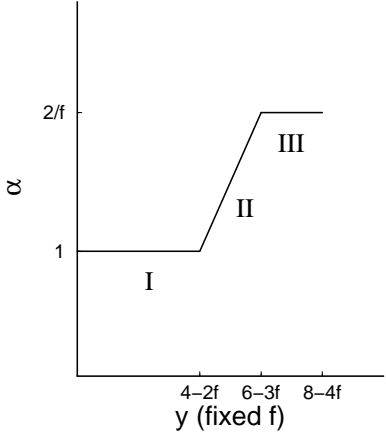}
\caption{Decay exponent for the $A+A \to \emptyset$ reaction in the
L\'{e}vy flight system with disorder model I in two dimensions:
 $c_A(t) \sim ({\rm
const}) t^{-\alpha}$.}
\label{fig4}
\end{figure}

In the region of $\epsilon > 0$, $\gamma_R^* = 2\pi\epsilon$ is the
nontrivial fixed point. But depending on the value of $\delta =
3f+y - 6$, the matching procedure yields different results. For
$\delta < 0$, i.e. in region II of Fig.~\ref{fig3}, there is no
non-trivial fixed point for $\lambda_R$. The corresponding asymptotic
concentration decay is a little faster than that in region I:
\begin{equation}
c_A(t) \sim \frac{1}{t_0} \left(\frac{1}{\lambda} + \frac{1}
{4\pi|\delta|D_f \Lambda^{f-2}} \right) \left(\frac{t}{t_0}\right)
^{-\frac{2-|\delta|}{f}}
\end{equation}
For $\delta > 0$, i.e. in region III of Fig.~\ref{fig3}, $\lambda_R^* =
4\pi\delta$ is the fixed point. The asymptotic concentration decay is
the fastest:
\begin{equation}
c_A(t) \sim \frac{1}{4\pi\delta D_f \Lambda^{f-2} t_0}
\left(\frac{t}{t_0}\right)^{-\frac{2}{f}}
\end{equation}
The relationship between the concentration decay exponent, $\alpha$,
and disorder range, $y$, at a fixed value of $f$ is plotted in
Fig~\ref{fig4}.  Note that the strength of the disorder has no effect
whatsoever on the concentration decay.

The reader will note that the disorder can never slow down the reaction
in this model. This is quite an unexpected result, as these reactions
are expected to become transport-limited at long times, and disorder
should slow down the transport. Model I, Eq.~(\ref{Fokker}), is
somewhat unphysical in that this cannot happen due to the lack of
disorder contribution to $Z_f$.

\section{Reaction in a Model of Disorder that Obeys Detailed Balance}
Although Eq.~(\ref{Fokker}) is often used in the literature, it does
not guarantee a long-time Boltzmann distribution for $G({\bf r},
t)$. That is, this form of the disorder does not satisfy detailed
balance. A more natural form of the Fokker-Planck equation for L\'{e}vy
flights in random disorder is
\begin{equation}
\frac{\partial G({\bf r}, t)}{\partial t} = \nabla^{f-1} \cdot [D_f
\nabla G({\bf r}, t) + G({\bf r}, t) \nabla V({\bf
r})]
\label{17}
\end{equation}
where $\nabla^{f-1}$ is the inverse Fourier transform of $-ik^{f-2} {\bf k}$.
Equation (\ref{17}) can be interpreted
as a modification of the continuity equation to take into account the
long-range transport induced by the L{\'e}vy flights;
it is derived in the appendix.
With the same form of the correlation function for the potential,
Eq.~(\ref{VV}), we have
\begin{eqnarray}
S &=& \int d^d {\bf x} \int_0^{t_f} dt \bar{a}_i({\bf x}, t) [\partial_t
- D_f(\nabla^2)^{f/2} + \delta(t)] a_i({\bf x}, t) \nonumber \\
  & & + \frac{\lambda}{2} \int d^d {\bf x} \int_0^{t_f} dt [2
\bar{a}_i({\bf x}, t) a_i^2({\bf x}, t) + \bar{a}_i^2({\bf x}, t)
a_i^2({\bf x}, t)] \nonumber\\
  & &- n_0 \int d^d {\bf x} \bar{a}_i({\bf x}, 0) 
\nonumber\\
&& - \frac{\gamma}{2}
\int dt_1 dt_2 \int_{{\bf k}_1 {\bf k}_2 {\bf 
  k}_3{\bf k}_4} (2\pi)^d  
  \delta({\bf k}_1 + {\bf k}_2 +  {\bf k}_3 + {\bf
k}_4)
\nonumber \\
&& \times
\frac{k_1^{f-2} k_3^{f-2} {\bf k}_1 \cdot ({\bf k}_1 + {\bf k}_2) {\bf
k}_3 \cdot ({\bf k}_3 + {\bf k}_4)}{|{\bf k}_1 + {\bf k}_2|^{2+y}}
\nonumber \\
&& \times
\hat{\bar{a}}_i ({\bf k}_1, t_1) \hat{a}_i ({\bf k}_2, t_1)
\hat{\bar{a}}_j ({\bf k}_3, t_2) 
    \hat{a}_j ({\bf k}_4, t_2) 
\end{eqnarray}

Again, to apply the field theoretic renormalization procedure, the
action is recast as
\begin{eqnarray}
S &=& \int d^d {\bf x} \int_0^{t_f} dt \bar{a}_i({\bf x}, t) [\partial_t
- Z_fD_{fR}(\nabla^2)^{f/2} \nonumber \\
  & &+ \delta(t)] a_i({\bf x}, t) + \frac{1}{2} \mu^{f-d} Z_{\lambda}
\lambda_R D_{fR} \int d^d {\bf x} \int_0^{t_f} dt \nonumber \\
  & & \times [2 \bar{a}_i({\bf x}, t) a^2_i({\bf x}, t) + \bar{a}_i^2({\bf
x}, t) a_i^2({\bf x}, t)] \nonumber \\
  & & - \mu^d n_{0R} \int d^d {\bf x} \bar{a}_i({\bf x}, 0) -
\frac{1}{2} \mu^{\epsilon} Z_{\gamma} \gamma_{R} D_{fR}^2 \int dt_1 dt_2
\nonumber \\
  & &\times \int_{{\bf k}_1 {\bf k}_2 {\bf k}_3{\bf k}_4} (2\pi)^d
\delta({\bf k}_1 + {\bf k}_2 +  {\bf k}_3 + {\bf k}_4) \nonumber \\
  & &
\times \frac{k_1^{f-2} k_3^{f-2} {\bf k}_1 \cdot ({\bf
k}_1 + {\bf k}_2) {\bf k}_3 \cdot ({\bf k}_3 + {\bf k}_4)}{|{\bf k}_1
+ {\bf k}_2|^{2+y}} 
\nonumber \\
&& \times
   \hat{\bar{a}}_i ({\bf k}_1, t_1) \hat{a}_i ({\bf k}_2, t_1)
  \hat{\bar{a}}_j ({\bf k}_3, t_2) \hat{a}_j ({\bf k}_4,
  t_2)\nonumber \\
\end{eqnarray}
where, $d_c = 2 + y$, $\epsilon = d_c - d$, and the rest of the
parameters are the same as in action I.

The connections between the renormalized and unrenormalized parameters
are
\begin{eqnarray}
Z_fD_{fR} &=& D_f \nonumber \\
\mu^{f-d} Z_{\lambda} \lambda_{R} D_{fR} &=& \lambda \nonumber\\
\mu^d n_{0R} &=& n_0 \nonumber\\
\mu^{\epsilon} Z_{\gamma} \gamma_{R} D_{fR}^2 &=& \gamma
\end{eqnarray}

The diagrams are the same as those in Figs.~\ref{fig1} and
\ref{fig2}, except that the diagrams from normal diffusion,
Figs.\  \ref{fig1}b and \ref{fig2}b,
are not present. 
A one-loop diagrammatic calculation gives
\begin{eqnarray}
Z_f &=& 1 + \frac{\gamma_R}{4\pi \epsilon}\nonumber\\
Z_{\lambda} &=& 1 - \frac{\gamma_R}{2\pi \epsilon} +
\frac{\lambda_R}{4\pi(f-d)}\nonumber\\
Z_{\gamma} &=& 1 + \frac{\gamma_R}{2\pi \epsilon}
\end{eqnarray}
where a double-pole expansion of $1/\epsilon$ and $1/(f-d)$ is used in
the calculation of $Z_{\lambda}$.

The dynamic exponent is given by
\begin{equation}
z = f - \mu \frac{\partial}{\partial \mu} \ln Z_f = f +
\frac{\gamma_R} {4\pi}
\end{equation}
This suggests that the L\'{e}vy flights are significantly slowed by the
presence of the new random disorder term. That is, when detailed
balance is obeyed, the disorder affects the dynamical exponent.

The $\beta$ functions give us the flow equations from $Z_i$
in two dimensions as 
\begin{eqnarray}
\frac{d\lambda_{R}}{dl} &=& -\beta_{\lambda_R} = (f-2)\lambda_R
+ \frac{3\gamma_R \lambda_R} {4\pi} - \frac{\lambda_R^2} {4\pi} \nonumber\\
\frac{d\gamma_R}{dl} &=& -\beta_{\gamma_R} = \epsilon \gamma_R
\end{eqnarray}
This action is well-behaved, and there is no regularization
difficulty. And indeed, both momentum-shell renormalization and
field-theoretic renormalization yield identical flow equations to one-loop
order. From the flow equations, for $\epsilon > 0$, $\gamma_R$ flows
to $\infty$, indicating that the disorder dominates over the L\'{e}vy
flights. For $\epsilon < 0$, $\gamma_R$ flows to $0$, indicating
transport of L\'{e}vy flights dominates over the disorder. This same
behavior, that the disorder must be adjusted to be compatible with the
transport, was found in the turbulent reactive flow
problem~\cite{Deem99}. An interesting system arises when those two
effects are competitive, and so we require $\epsilon = 0$, i.e. $y =
0$. Under this constraint, $\gamma_R = \gamma/D_f^2$ does not
flow. Very likely, this result holds to all orders. Further analysis
of the flow equation for $\lambda_R$ indicates two regimes. For
$\gamma/D_f^2> 4(2-f)\pi/3$, or strong disorder, $\lambda_R$ has a
stable nontrivial fixed point, given by
\begin{equation}
\lambda_R^* = 4\pi(f - 2)+\frac{3\gamma}{D_f^2} = \lambda^* \Lambda^{2-f}/D_f
\end{equation}
Following the matching procedure, we have
\begin{equation}
c_A(t) \sim \frac{1}{\lambda^* t} \left(\frac{t}{t_0}
\right)^{-\frac{2}{f+\gamma/(4\pi D_f^2)} + 1}
\end{equation}
However, for $\gamma/D_f^2< 4(2-f)\pi/3$, or weak disorder, there is no
nontrivial fixed point for $\lambda_R$, and we have
\begin{eqnarray}
c_A(t) &\sim & \left(\frac{1}{\lambda} + \frac{1} {[4\pi(2-f)-3\gamma /
D_f^2] \Lambda^{f-2} D_f} \right)\nonumber \\ 
& & \times \frac{1}{t} \left(\frac{t}{t_0}\right) ^{-\frac{\gamma
/(2\pi D_f^2)}{f+\gamma /(4\pi D_f^2)}}
\end{eqnarray}
For the special case of $\gamma/D_f^2= 4(2-f)\pi/3$, $\lambda_R$ decays
marginally, and we have a logarithmic correction to the decay:
\begin{equation}
c_A(t) \sim \frac{3\ln(t/t_0)}{8\pi D_f\Lambda^{f-2}(1+f) t}
\left(\frac{t}{t_0} \right)^{-\frac{2-f}{1+f}}
\end{equation}

\begin{figure}[t]
\centering
\leavevmode
\includegraphics[height=2in, angle=-90]{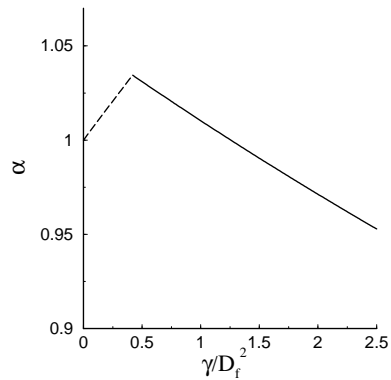}
\caption{Decay exponent for the $A+A \to \emptyset$ reaction in the
L\'{e}vy flight system that obeys detailed balance in two
dimensions: $c_A(t) \sim ({\rm
const}) t^{-\alpha}$. The figure is shown for $f=1.9$. The reaction is
transport limited on the solid curve and reaction limited on the
dashed curve. Note that the region $\alpha>1$ corresponds to
'superfast' reaction.}
\label{fig5}
\end{figure}

In the present case, unlike with the action of section II, the
range and strength of disorder affect the decay exponent significantly. The
relationship between the decay exponent and $\gamma$ is plotted in
Fig.~\ref{fig5}. We see that a small amount of potential disorder
leads to an increased rate of reaction in the L\'{e}vy flights. But as
the potential disorder increases further, the rate of reaction
eventually decreases. This figure is very much similar to the one that
showed up in the study of reactive turbulent flow~\cite{Park98}. In
fact, if the L\'{e}vy flight parameters are related to the turbulence
parameters~\cite{Park98} as $f=2-g^*_{\sigma}$ and $\gamma = 4\pi D_f^2
g^*_{\gamma}$, these two models predict {\it exactly} the same concentration
decays. As suggested in~\cite{Park98}, in order for the reaction to
occur, multiple reactants must be trapped in regions of low potential
energy. After the trapped reactants are depleted, new reactants must
be replenished by rapid transport to continue the reaction. Certain
combinations of fast transport, $f<2$, and disorder, $\gamma$, lead to
'superfast' reaction, $\alpha>1$, as shown in
Fig.~\ref{fig5}. Interestingly, this result means that the inhomogeneous
system can have a faster reaction rate than the homogeneous,
well-mixed system.

\section{Conclusion}
We have analyzed the $A+A \to \emptyset$ reaction in two dimensional
L\'{e}vy flight systems using two models of random disorder.
 For a common model in the literature, the dynamic exponent
always locks to the microscopic step index $f$, and the reaction decay
exponent varies between 1 and $2/f$. This surprisingly unphysical
result that the disorder cannot slow down the reaction is due to the
fact that this model does not satisfy detailed balance. For a model
that does satisfy detailed balance, on the other hand, the disorder
can and does modify the transport properties of the system. When the
disorder is adjusted to be compatible with the L\'{e}vy flight
statistics, the reaction decay exponent first rises above unity and
then drops to zero as the strength of disorder is increased. These
results are identical to those from reactive turbulent flow, and this
harmony suggests that L\'{e}vy flights, properly interpreted, can be a
viable model of turbulent fluid transport.

\section*{Acknowledgment}
This research was supported by the Alfred P. Sloan Foundation through
a fellowship to M.W.D.

\section*{Appendix}
In this appendix, we derive Eq.~(\ref{17}) from a master equation.
Particle transport by L{\'e}vy flight statistics can be
modeled by hopping dynamics of particles in space.
The hopping rate in the absence of an external potential
is simply proportional to Eq.~(\ref{1}).  In the presence of
an external potential, we modify this rate in a way that
satisfies detailed balance \cite{victor}:
\begin{equation}
\alpha ({\bf r}_{\rm o} \to {\bf r}_{\rm n}) =
    (D_f/r_0^f) P({\bf r}_{\rm n} -{\bf r}_{\rm o}) 
  e^{-\beta [V({\bf r}_{\rm n}) -V({\bf r}_{\rm o})]/2}
\end{equation}
where $\alpha({\bf r}_{\rm o} \to {\bf r}_{\rm n})$ is the rate
at which particles hop from ${\bf r}_{\rm o}$ to ${\bf r}_{\rm n}$.
The master equation for the Green's function of this
process is
\begin{eqnarray}
\frac{\partial G({\bf r})}{\partial t} &=& 
\int d^d {\bf x} ~ \alpha({\bf r} - {\bf x} 
\to {\bf r}) G({\bf r} - {\bf x}) 
\nonumber \\
&& -
\int d^d {\bf x} ~ \alpha({\bf r} 
\to {\bf r} + {\bf x}) G({\bf r})
\end{eqnarray}
We now expand this master equation to first order in $\beta$,
\emph{i.e.}\ we look for a Fokker-Planck equation that is linear in the
potential.
Noting that $\hat P({\bf k}) =1  -r_0^f k^f$, we find
\begin{eqnarray}
\frac{\partial \hat G({\bf k})}{\partial t} &=& 
-D_f k^f \hat G({\bf k}) 
 - 
\frac{\beta D_f}{2} \int_{\bf h} \hat G({\bf k} - {\bf h}) \hat V({\bf h})
\nonumber \\ 
&&\times [k^f + h^f - \vert {\bf k} - {\bf h} \vert^f] + O({\beta^2}) 
\label{25}
\end{eqnarray}
In real space, we would write this equation as 
\begin{eqnarray}
\frac{\partial G({\bf r})}{\partial t} = &&
D_f (\nabla^2)^{f/2} G 
+ 
\frac{\beta D_f}{2} \bigg[(\nabla^2)^{f/2}(G V) 
\nonumber \\ 
&&+
G (\nabla^2)^{f/2} V - V 
(\nabla^2)^{f/2} G \bigg]
\end{eqnarray}
Since we are interested in how the long-wavelength features of the
potential affect the dynamics, \emph{i.e.}\ we take into account
only the small wavevector
contributions in Eq.~(\ref{VV}), we make an expansion 
of Eq.~(\ref{25}) for small $h$.
We find 
$(k^f + h^f - \vert {\bf k} - {\bf h} \vert^f)/2 \sim 
f k^{f-2} {\bf k} \cdot {\bf h}/2$
for $1 < f \le 2$.
 Realizing that
the Green's function should relax to the Boltzmann distribution
at long times, and that
$D_f$ and $\beta$ may be renormalized a finite amount
by a variety of irrelevant operators, we conclude that the proper
long-wavelength theory should be
\begin{eqnarray}
\frac{\partial \hat G({\bf k})}{\partial t} &=& 
-D_f k^f \hat G({\bf k}) 
\nonumber \\ 
&& - 
\beta D_f \int_{\bf h} \hat G({\bf k} - {\bf h}) \hat V({\bf h})
k^{f-2} {\bf k} \cdot {\bf h} 
\end{eqnarray}
which is exactly Eq.~(\ref{17}) with units of $\beta$
inserted.

\bibliography{levy}

\end{document}